\documentstyle[prl,floats,aps,twocolumn,epsf,graphicx]{revtex}
\begin{document}
\twocolumn[\hsize\textwidth\columnwidth\hsize\csname
@twocolumnfalse\endcsname

\title{Phase-Transition in Binary Sequences with Long-Range Correlations}
\author{Shahar Hod$^{1,2}$ and Uri Keshet$^2$}
\address{$^1$The Racah Institute of Physics, The Hebrew University, Jerusalem 91904, Israel}
\address{}
\address{$^2$Department of Condensed Matter Physics, Weizmann Institute, Rehovot
 76100, Israel}
\date{\today}
\maketitle

\begin{abstract}

\ \ \ Motivated by novel results 
in the theory of correlated sequences, we analyze the dynamics of random walks with long-term 
memory (binary chains with long-range correlations). 
In our model, the probability for a unit bit in a binary string depends on the 
{\it fraction} of unities preceding it. 
We show that the system undergoes a dynamical phase-transition from normal 
diffusion, in which the variance $D_L$ scales as the string's length $L$, 
into a super-diffusion phase ($D_L \sim L^{1+|\alpha|}$), when the correlation strength 
exceeds a critical value.
We demonstrate the generality of our results with respect to alternative models, and discuss 
their applicability to various data, such as 
coarse-grained DNA sequences, written texts, and financial data. 
\end{abstract}
\bigskip

]

Dynamical systems with long-range spatial (and/or temporal) correlations 
are attracting considerable interest across many disciplines. 
They are identified in physical, biological, social, and economic sciences 
(see e.g., [1-6] and references therein). 
Of particular interest are situations 
in which the system can be mapped onto a mathematical object, such as 
a correlated sequence of symbols, preserving the essential statistical properties 
of the original system. 

One of the methods most frequently used to obtain insight into the nature of correlations in 
a dynamical system consists of mapping the space of states onto two symbols \cite{Usa}. 
Thus, the problem is reduced to the exploration of the statistical properties of correlated 
binary chains. 
This can also be viewed as the analysis of a history-dependent random walk. 
Random walk is one of the most ubiquitous concepts of statistical physics. It lends 
applications to numerous scientific fields (see e.g., \cite{BaNi,Kam,FeFrSo,Wei,AvHa,DiDa,Hod} and 
references therein). 

It is well established that the statistical properties of 
coarse-grained DNA strings and written texts significantly deviate from those of purely 
random sequences \cite{Kan,Sch}. Financial data (such as stock market quotes) are similarly 
far from being pure-diffusive. Moreover, these systems exhibit 
``super-diffusive'' behavior in the sense that the variance $D(L)$ grows asymptotically {\it faster} than $L$ 
(where $L$ is the length of the considered text). Specifically, 
$D \sim L^{\alpha}$, with $\alpha > 1$ \cite{Usa}. 
Such a remarkable (and essentially universal) phenomenon can be attributed 
to long-range positive correlations. 
Systems with such correlations may be anticipated to exhibit a dynamical phase transition 
(from normal to super diffusive behavior) at some critical correlation strength. 

Thus, the problem of random walk where the jumping probabilities are history-dependent is 
of great interest for understanding the behavior of systems with long-range correlations, such 
as DNA strings, written texts, and financial data. 
The aim of the present Letter is to analyze this problem, and to provide a simple yet generic 
{\it analytical} description of the statistical properties of these systems.

We begin by solving a simple model which incorporates long-range correlations into an otherwise random 
sequence. We consider a discrete binary string of symbols, $a_i=\{0,1\}$, in which 
the conditional probability of a given symbol (say, a unit bit) occurring at the position $L+1$ 
is {\it history-dependent}, and given by 

\begin{equation}\label{Eq1}
p(k,L)={1 \over 2}\Big(1-\mu {{L-2k} \over {L+L_0}}\Big)\  ,
\end{equation}
where $k$ is the number of such symbols (unities) appearing in the preceding $L$ bits. 
The correlation parameter $\mu$, where $-1< \mu < 1$, 
determines the strength of correlations in the system. 
The persistence condition $\mu>0$ implies that a given symbol 
in the preceding sequence promotes the birth of a new identical symbol. 
On the other hand, in the anti-persistence region $\mu < 0$, each 
symbol inhibits the appearance of a new identical symbol. 
The parameter $L_0>0$ is a constant transient time. For $L \ll L_0$ the sequence is approximately 
random (uncorrelated), whereas for $L \gg L_0$ the effect of correlations takes over \cite{Note1}. 

In this model, the conditional probability $p(k,L;\mu,L_0)$ depends on the {\it fraction} 
of unities (or zeroes) in the preceding bits, and is independent of their arrangement. 
This allows one to obtain an {\it analytical} description of the system's dynamical behavior. 
As we shall demonstrate below, this simple model provides a good quantitative 
description of the observed statistical properties of various natural systems, such as 
coarse-grained DNA strings, written texts, and financial data.

The probability $P(k,L+1)$ of finding $k$ identical symbols (say, unities) 
in a sequence of length $L+1$ follows the evolution equation

\begin{eqnarray}\label{Eq2}
P(k,L+1) & = &[1-p(k,L)]P(k,L) \nonumber \\
&& +p(k-1,L)P(k-1,L)\  .
\end{eqnarray} 
Crossing to the continuous limit, one obtains the 
Fokker-Planck diffusion equation for the correlated process

\begin{equation}\label{Eq3}
{{\partial P} \over {\partial L}}={1 \over 2} {{{\partial^2 P} \over {\partial x}^2}}
-{{\mu} \over {L+L_0}}{{\partial(xP)} \over \partial x}\  ,
\end{equation}
where $x \equiv 2k-L$. The evolution equation (\ref{Eq3}) along with the 
initial condition $P(x,t=0)=\delta(x)$, has a solution in the 
form of a Gaussian distribution

\begin{equation}\label{Eq4}
P(x,L)={1 \over {\sqrt{{2\pi D(L)}}}} \exp\Big[-{{x^2} \over {2D(L)}}\Big]\  ,
\end{equation}
where the variance $D(L)$ is given by

\begin{equation}\label{Eq5}
D(L;\mu,L_0)={{L+L_0} \over {1-2\mu}} \Big[ 1 -{\Big({{L_0} \over {L+L_0}}\Big)}^{1-2\mu}\Big]\  .
\end{equation} 
Equation (\ref{Eq5}) 
breaks down at the special case $\mu= {1 \over 2}$, in which case the variance is given by 

\begin{equation}\label{Eq6}
D(L;\mu_c,L_0)=(L+L_0)\ln\Big({{L+L_0} \over {L_0}}\Big)\  .
\end{equation}

Remarkably, one finds that the correlated system undergoes a dynamical phase transition 
at the critical correlation strength $\mu_c \equiv {1 \over 2}$. 
The variance $D(L)$ of the correlated sequence has three qualitatively different 
asymptotic behaviors (in the $L \gg L_0$ limit)

\begin{equation}\label{Eq7}
D(L) \simeq \cases{ 
(1-2\mu)^{-1}L & $\mu<\mu_c$\  ; \cr
L \ln (L/L_0) & $\mu=\mu_c$\  ; \cr
(2\mu-1)^{-1}{L_0}^{1-2\mu}L^{2\mu} & $\mu>\mu_c$\  . \cr }
\end{equation}
Thus, for $\mu < \mu_c$ the asymptotic variance scales linearly with the string length, 
whereas for a history-dependent chain with strong positive correlations ($\mu > \mu_c$) the system is 
characterized by a super-diffusion phase, in which case $D(L)$ grows asymptotically 
faster than $L$ \cite{Note2}.

The analytical model can readily be extended to encompass situations in which the binary sequence 
is {\it biased}. Let

\begin{equation}\label{Eq8}
p(k,L)={1 \over 2}\Big(1+q-\mu {{L-2k} \over {L+L_0}}\Big)\  ,
\end{equation}
with $-1<q<1$. The distribution $P(x,L)$ corresponding to this conditional 
probability is given by a Gaussian function, centered about the position 

\begin{equation}\label{Eq9}
x_c(L)={{q} \over {1-\mu({{L} \over {L+L_0}})}}L\  .
\end{equation}
Thus, the drift velocity approaches an asymptotically constant value ${{q} \over {1-\mu}}$. 
The variance $D(L)$, unaltered by the bias is given by Eqs. (\ref{Eq5}) and (\ref{Eq6}).

\begin{figure}[tbh]
\centerline{\epsfxsize=9cm \epsfbox{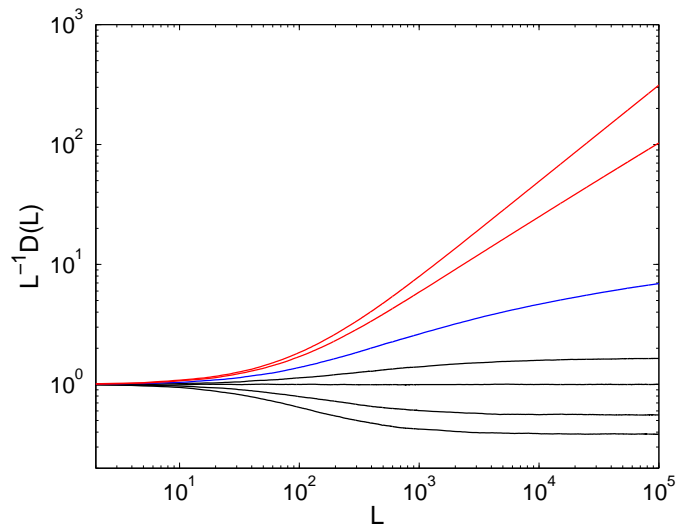}} 
\caption{The scaled variance $L^{-1}D(L)$ as a function of the string length $L$. 
We present results for $\mu=-0.8, -0.4, 0, 0.2, 0.5, 0.8$, and $0.9$ (from bottom to top), 
with $L_0=100$. 
The numerically computed asymptotic slopes agree with the analytical predictions [see Eqs. 
(\ref{Eq5}) and (\ref{Eq6})] to within less than $1\%$.}
\label{Fig1}
\end{figure}

In order to confirm the analytical results, we perform numerical simulations of (discrete) 
binary sequences. Figure \ref{Fig1} displays the resulting scaled variance $L^{-1}D(L)$ of 
correlated strings with various different values of the correlation parameter $\mu$. 
We find an excellent agreement between the analytically predicted results [see Eqs. (\ref{Eq5}) and 
(\ref{Eq6})] and the numerical ones.

\begin{figure}[tbh]
\centerline{\epsfxsize=9cm \epsfbox{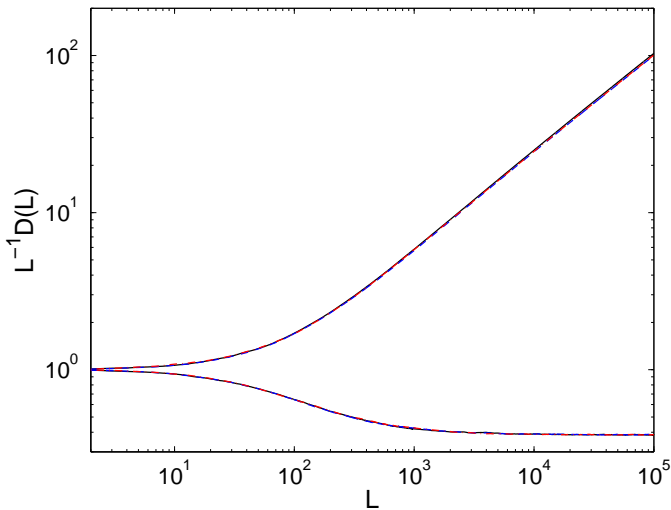}} 
\caption{The scaled variance $L^{-1}D(L)$ for three different forms of the 
function $F(\xi)$: $\xi$, ${2 \over \pi}\sin({\pi \over 2}\xi)$, and $\tanh(\xi)$. 
We present results for $\mu=-0.8$ and $\mu=0.8$,with $L_0=100$. The different curves are 
almost indistinguishable.} 
\label{Fig2}
\end{figure}

{\it Robustness of the linear model.--} 
In order to show the generality of the model discussed above, we consider situations in which 
the (history-dependent) jump probability is an arbitrary odd function \cite{Note3} of 
the fraction $\xi \equiv {x \over {L+L_0}}$ of unities (zeroes) that appeared in the previous $L$ symbols

\begin{equation}\label{Eq10}
p(x,L)={1 \over 2}[1+\mu F(\xi)]\  .
\end{equation}
For asymptotically large $L$, one always finds $\xi\to 0$ for non-ballistic diffusion, 
justifying a power-law expansion of $F(\xi)$. 
As long as this expansion includes a linear term, the original differential equation (\ref{Eq3}) 
is recovered for large $L$. We therefore expect the previous analytical results 
[Eqs. (\ref{Eq5}) and (\ref{Eq6})] to hold true for generic ({\it non}-linear) models as well. 
The generality of the model is illustrated in Fig. \ref{Fig2}, in which we depicts 
results for various choices of the probability function $F(\xi)$. As predicted, the results 
are found to agree with the linear model.

{\it Applications.--} 
The robustness of the linear model (see Fig. \ref{Fig2}) suggests 
that it may capture the essence of the 
underlying correlations in a diversity of systems in nature. 
We therefore examine the use of the results derived in the present work as an analytical explanation for the 
observed statistical properties of natural systems, such as 
DNA strings, written texts, and financial data.

As mentioned, it is well established that these systems often exhibit a significant 
deviation from random sequences \cite{Kan,Sch}, and are characterized by a 
``super-diffusive'' behavior in which $D \sim L^{\alpha}$, with $\alpha > 1$ \cite{Usa}. In such 
systems, super-diffusion may be attributed to long-range (positive) correlations. In fact, 
the analytical model allows one to determine the correlation strength of these chains. 

\begin{figure}[tbh]
\centerline{\epsfxsize=9cm \epsfbox{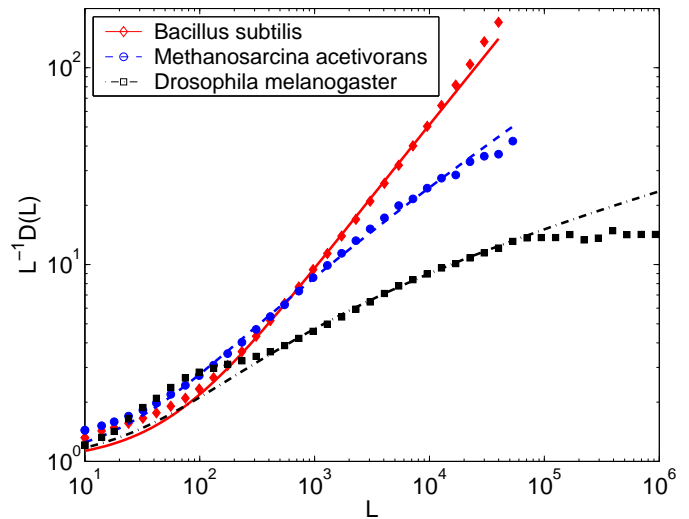}} 
\caption{The scaled variance $L^{-1}D(L)$ as a function of the string length $L$, 
for coarse-grained DNA sequences of various organisms.
The mapping and parameters used are given in Table I. 
Theoretical results [see Eq. (\ref{Eq5})] are represented by curves.}
\label{Fig3}
\end{figure}

Figure \ref{Fig3} depicts the scaled variance $L^{-1}D(L)$ calculated from DNA sequences of 
various organisms, as a function of the string length $L$. 
It is of considerable interest to examine in such methods the statistical 
properties characterizing the DNA of organisms in various evolutionary levels: 
Bacillus subtilis ({\it Bacteria}), Methanosarcina acetivorans ({\it Archaea}), 
and Drosophila melanogaster ({\it Eukarya}) \cite{Usa,DNAs}. 
The theoretical model provides a good description of the 
empirical data \cite{Note4}, attributing different correlation strengths $\mu$ to different organisms, as 
summarized in Table \ref{Tab1}. 

The super-diffusive behavior, shown in Fig. \ref{Fig3} to persist across very long sequences is highly suggestive 
of {\it long}-range correlation extending over {\it more} than one 
gene (e.g., $\sim 5 \times 10^4$ base-pairs in Drosophila).

Next, we have applied the results of the analytical model to various coarse-grained written
texts \cite{Kan,Sch,Usa}. It has long been recognized that the corresponding binary strings are highly 
self-correlated. The present analytical model enables one to determine quantitatively the strength of these 
inner correlations; see Table \ref{Tab1}.

\begin{table}
\caption{The correlation strength parameter $\mu$ for various binary strings. 
We use the following mappings: $\{A,G\} \to 0$, $\{C,T\} \to 1$ for DNA sequences [5,18]; 
(a to m) $\to 0$, (n to z) $\to 1$ for written texts [5]; and daily fall $\to 0$, daily rise 
$\to 1$ for stock market quotes [20].}
\label{Tab1}
\begin{tabular}{llc}
Data Type & String Source & $\mu$ \\
\tableline
DNA sequences 
& Drosophila melanogaster & $0.57$ \\
& Methanosarcina acetivorans & $0.70$ \\
& Bacillus subtilis & $0.86$\\
Written texts 
& Alice's adventures in wonderland& $0.58$ \\
& The Holy Bible in English & $0.84$ \\
& Works on computer science & $0.88$ \\
Stock markets
& NASDAQ & 0.39 \\
& DJIA & 0.76 \\
\end{tabular}
\end{table}

\begin{figure}[tbh]
\centerline{\epsfxsize=9cm \epsfbox{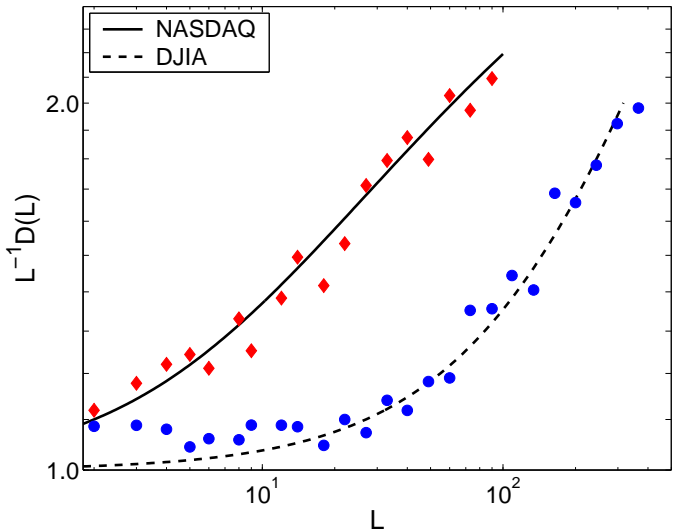}} 
\caption{The scaled variance $L^{-1}D(L)$ as a function of the sequence length $L$, 
for coarse-grained financial data: DJIA and NASDAQ daily quotes [20]. 
The mapping and parameters used are given in Table I. Theoretical results 
[see Eq. (\ref{Eq5})] are represented by curves.}
\label{Fig4}
\end{figure}

In Figure \ref{Fig4} we show the scaled variance of coarse-grained 
financial data (daily quotes of the Dow Jones Industrial Average, and the NASDAQ \cite{Djia}). 
We note that the linear model underestimates the 
empirical variance at {\it short} time scales. This fact can be traced back to 
short-term correlations in the markets. (It is interesting to note that the DJIA maintains 
an approximately normal diffusive behavior for a period of about one month). 
However, this short-term memory is washed out at longer time 
scales, in which case the analytical model provides a good description of the 
empirical results, as evident from Fig. \ref{Fig4}. The corresponding values of the 
correlation parameter $\mu$ are summarized in Table \ref{Tab1}.

In summary, in this Letter we have analyzed the dynamics of random walks with
{\it history-dependent} jump probabilities. 
Our work was motivated not only by the intrinsic interest in such dynamical
processes, but also by the flurry of activity in the field of long-range
correlated systems, and by some universal statistical features observed in many 
different natural systems.

We have broadened the study of binary strings to include long-range
correlations, extending throughout the length of the chain.
Using a simple and exactly solvable model, we identify a dynamical phase
transition, from normal diffusion [$D(L) \sim L$] to super-diffusive
behavior [$D(L) \sim L^{2 \mu}$], taking place as the correlation parameter $\mu$
exceeds its critical value. 
We show that in spite of the simplicity of the model, it is robust, and can
easily be extended to describe various features (such as a biased history-dependent random 
walk or sub-diffusion).

Next, we have applied the analytical results of the model to various binary strings, extracted 
from very different natural systems, such as 
coarse-grained DNA sequences, written texts, and financial data. 
We find that the model adequately describes the long-term behavior of these systems. 
Furthermore, the model provides a straightforward method to measure the
correlation strength of these systems. 
Our results can be applied to various natural systems, and may shed light on the
underlying rules governing their dynamics. 
For example, the super-diffusive behavior of DNA sequences (see Fig. \ref{Fig3}) suggests 
long-range correlations extending across more than one gene. The model attributes 
different correlation strengths to different organisms.
 
\bigskip
\noindent
{\bf ACKNOWLEDGMENTS}
\bigskip

SH thanks a support by the Dr. Robert G. Picard fund in physics. 
We would like to thank Oded Agam, Yitzhak Pilpel, Eli Keshet, Ilana Keshet, 
Clovis Hopman, Eros Mariani, Assaf Pe`er, Oded Hod, and Ehud Nakar for helpful discussions. 
We thank O. V. Usatenko and V. A. Yampol`skii for providing us with their data. 
This research was supported by grant 159/99-3 from the Israel Science Foundation.


\begin{thebibliography}{99}
\bibitem{DNAs} DNA sequences of various organisms were obtained from ftp://ftp.ncbi.nih.gov/genomes.
\end{thebibliography}
\end{document}